\documentclass[notitlepage]{article}
\date{today}
\usepackage{natbib,hyperref}
\usepackage{graphicx}
\usepackage{epstopdf}
\usepackage{amsmath}
\usepackage[final]{changes}
\usepackage{authblk}

\newcommand{\ket}[1]{\left|#1\right>}

\newcommand{\beq}{\begin{equation}}
\newcommand{\eeq}{\end{equation}}
\newcommand{\beqa}{\begin{eqnarray}}
\newcommand{\eeqa}{\end{eqnarray}}

\begin{document}
\title{Progress on exchange interaction induced spin manipulation and decoherence in quantum dots} 
\title{Progress on spin decoherence in quantum dots for quantum computing applications} 
\title{Exchange interaction induced spin manipulation and decoherence in quantum dots for quantum computing applications} 
\title{Progress on manipulation and dephasing of two-electron spins in quantum dots for spin-based quantum computing}
\title{Progress on dephasing and manipulation of two-electron spins in quantum dots for spin-based quantum computing}
\title{Spin dephasing in quantum dots for spin-based quantum computing}
\title{{Dephasing of Exchange-coupled Spins in Quantum Dots for Quantum Computing}}
\author{Peihao Huang\thanks{huangph@sustech.edu.cn}}
\affil{Shenzhen Institute for Quantum Science and Engineering, Southern University of Science and Technology, Shenzhen 518055, China}
\affil{Guangdong Provincial Key Laboratory of Quantum Science and Engineering, Southern University of Science and Technology, Shenzhen, 518055, China}


\date{\today}
\maketitle


\begin{abstract}


A spin qubit in semiconductor quantum dots holds promise for quantum information processing for scalability and long coherence time. An important semiconductor qubit system is a double quantum dot trapping two electrons or holes, whose spin states encode either a singlet-triplet qubit or two single-spin qubits coupled by exchange interaction. In this article, we report progress on spin dephasing of two exchange-coupled spins in a double quantum dot. We first discuss the schemes of two-qubit gates and qubit encodings in gate-defined quantum dots or donor atoms based on the exchange interaction. Then, we report the progress on spin dephasing of a singlet-triplet qubit or a two-qubit gate. The methods of suppressing spin dephasing are further discussed. The understanding of spin dephasing may provide insights into the realization of high-fidelity quantum gates for spin-based quantum computing.
\end{abstract}

\section{Introduction}

A semiconductor spin qubit is promising for quantum computing because of the long spin coherence time and the compatibility with the technological standards in the semiconductor industry \cite{loss1998, kane1998, petta2005, hanson_spins_2007, morton_embracing_2011, zwanenburg_silicon_2013, zhang_semiconductor_2019, burkard_superconductorsemiconductor_2020, clerk_hybrid_2020, chatterjee_semiconductor_2021}.
A semiconductor spin qubit is defined by the spin states of an electron, {hole, or nuclei} in a semiconductor quantum dot (QD) \cite{loss1998, kane1998}, where the QD is an artificial structure that provides electronic confinement in all directions. There are different kinds of QDs, such as gate-defined QDs or dopant-based QDs, {for spin-based quantum computing} \cite{hanson_spins_2007, zwanenburg_silicon_2013}. {There are also self-assembled QDs important for applications in quantum photonics \cite{vamivakas_observation_2010, atature_material_2018, hepp_semiconductor_2019, cao_high-fidelity_2019, zhao_advanced_2020}.}  A gate-defined QD traps electrons from a two-dimensional electron gas (2DEG) in a nano-scale regime with electric potential from metallic gates on top of semiconductors \cite{hanson_spins_2007}. Whereas, a dopant-based QD traps electrons in a single dopant or a dopant cluster \cite{zwanenburg_silicon_2013}. Spin qubits in QDs carry quantum information, which is processed by applying electrical and/or magnetic fields \cite{skinner_hydrogenic_2003, taylor_fault-tolerant_2005, hill_surface_2015}.

An important spin-qubit system is a double quantum dot (DQD) with two electrons, where the spin states of electrons encode either a singlet-triplet \added{($S-T_0$)} qubit or two coupled single-spin qubits \cite{petta2005}.
The coupling of electron spins is through the exchange interaction arising from the Pauli principle in quantum mechanics for the indistinguishable elementary particles \cite{burkard_coupled_1999, hu_hilbert-space_2000}. 
The exchange interaction is one of the most important mechanisms for the realization of {two-qubit} quantum gates of semiconductor spin qubits {due to the strong coupling strength and the electrical tunability} \cite{loss1998, kane1998}.
For example, the controlled-phase (CPHASE) gate, {the controlled-rotation (CROT) gate}, and the SWAP gate based on the exchange interaction are realized in recent experiments \cite{veldhorst2015, watson_programmable_2018, zajac_resonantly_2018, huang_fidelity_2019, xue_benchmarking_2019, sigillito_coherent_2019, he_two-qubit_2019, hendrickx_fast_2020}. 
{Besides the exchange interaction, the two-qubit coupling is also possible by using the super-exchange interaction, the dipole-dipole interaction, or the strong coupling of spin qubits to a superconductor cavity \cite{anderson_antiferromagnetism_1950, friesen_efficient_2007, baart_coherent_2017, deng_interplay_2020, shulman_demonstration_2012, viennot_coherent_2015, tosi_silicon_2017, mi_coherent_2018, samkharadze_strong_2018, borjans_resonant_2020}. 
}

For the application in fault-tolerant quantum computing or the near-term intermediate-scale quantum, it is necessary to have a long coherence time of a qubit for high-fidelity quantum gates \cite{nielsen2010, preskill_quantum_2018}.
However, quantum information stored in qubits is fragile, and the qubit coherence tends to be destroyed by its coupling to environmental noise \cite{taylor2007}.
It is important to understand the underlying decoherence mechanisms to further extend the coherence time of a qubit for high-fidelity quantum gates \cite{tyryshkin_electron_2003, morton_solid-state_2008, simmons_entanglement_2011, tyryshkin_electron_2012, hu2006,bertrand_quantum_2015, reed_reduced_2016, martins_noise_2016, huang_spin_2018}. 
{Other aspects that are as important as decoherence are to increase the single-spin gate speed through electric dipole spin resonance (EDSR) \cite{rashba_orbital_2003, flindt_spin-orbit_2006, golovach2006, nowack_coherent_2007, rashba_theory_2008, tokura_coherent_2006, pioro-ladriere_micromagnets_2007, pioro-ladriere_electrically_2008, kawakami_electrical_2014, wu_two-axis_2014, yoneda2018}, improve robustness against noise through quantum control sequences or qubit encoding \cite{divincenzo_universal_2000, laird_coherent_2010, wang_composite_2012, zhang_randomized_2017, buterakos_exact_2021, medford_scaling_2012, medford_selfconsistent_2013, medford_quantumdotbased_2013, malinowski_symmetric_2017, shi_fast_2012, shi_fast_2014, kim_high-fidelity_2015, kim_microwave-driven_2015, cao_tunable_2016, shim_charge-noise-insensitive_2016, nichol_high-fidelity_2017, zhang_randomized_2017, throckmorton_fast_2017, russ_quadrupolar_2018, benito_electric-field_2019, friesen_decoherence-free_2017, yang_suppression_2017, shim_barrier_2018}, and design qubits for better scalability \cite{hu_strong_2012, petersson_circuit_2012, zajac_reconfigurable_2015, viennot_coherent_2015, tosi_silicon_2017, tosi_robust_2018, morello_donor_2020, mi_coherent_2018, samkharadze_strong_2018, borjans_resonant_2020, madzik_conditional_2021}. There are also theoretical studies of the quantum error correction scheme to suppress errors of spin qubits \cite{gross_hardwareefficient_2021}.
In the report, we focus on the decoherence mechanism of spin qubits given that it is the basis for many other aspects, such as quantum control, qubit encodings, etc.
}

{
Spin decoherence, including pure dephasing and relaxation, can arise from nuclear spins through hyperfine coupling \cite{cywinski2009, cywinski2009b, barnes2012, witzel2012}. It can also result from electric noises, such as phonon or charge noise, through the exchange interaction or spin-orbit coupling (SOC). 
Spin decoherence from nuclear spins is significantly suppressed in group IV materials with isotopic enrichment \cite{tyryshkin_electron_2003, morton_solid-state_2008, simmons_entanglement_2011, tyryshkin_electron_2012}. It is also generally slow for spin relaxation due to the electric noises through the SOC \cite{khaetskii2001, golovach2004, yang2013, tahan_relaxation_2014, huang_electron_2014, huang_spin_2014}. 
For example, the spin relaxation is about 10 $s^{-1}$ in silicon at a 2 T magnetic field \cite{amasha_electrical_2008, hayes_lifetime_2009, xiao_measurement_2010}. 
In a two-spin system in a DQD of our interest, the dominant mechanism of spin decoherence is the spin pure dephasing due to charge noise through the exchange interaction \cite{hu2006, bertrand_quantum_2015, reed_reduced_2016, martins_noise_2016, huang_spin_2018}. 
}

{
Different host materials of a DQD have been explored over the years.
Electron-spin qubits in a DQD were first realized in GaAs a decade ago, where the two-qubit gate, such as the SWAP gate or CROT gate, has been achieved between spin qubits \cite{petta2005, koppens_driven_2006}. 
Materials  with strong SOC, such as InAs, InSb, have also been explored for fast single-spin manipulations \cite{nadj-perge_spin-orbit_2010, petersson_circuit_2012, berg_fast_2013}. 
However, the nuclear spin noise in the host material of III-V semiconductor prohibits the further extension of the spin coherence time \cite{cywinski2009, cywinski2009b, barnes2012, witzel2012}. 
More recently, two-qubit gates are realized in group IV materials, such as silicon, with improved fidelities \cite{veldhorst2015, watson_programmable_2018, zajac_resonantly_2018, huang_fidelity_2019, xue_benchmarking_2019, sigillito_coherent_2019, he_two-qubit_2019, hendrickx_fast_2020}. Besides the electron system, a hole spin qubit in silicon or germanium is also under intensive study recently due to high mobility, suppressed coupling to nuclear noise, fast EDSR, and ease of fabrication \cite{fischer_spin_2008, watzinger_heavyhole_2016, watzinger_germanium_2018, li_coupling_2018, hendrickx_gatecontrolled_2018, hendrickx_fast_2020, scappucci_germanium_2020, 
xu_dipole_2020, gao_sitecontrolled_2020, hendrickx_fourqubit_2021, mutter_natural_2021, terrazos_theory_2021, zhang_anisotropic_2021, ono_hole_2017, nigg_superconducting_2017, kloeffel_direct_2018, crippa_electrical_2018, liles_spin_2018, venitucci_electrical_2018, venitucci_simple_2019, studenikin_electrically_2019, lawrie_spin_2020, kobayashi_engineering_2021, froning_strong_2021, froning_ultrafast_2021, camenzind_spin_2021, wang_optimal_2021, jirovec_singlettriplet_2021, bosco_hole_2021}.}


In this article, we report recent progress on the study of decoherence of {exchange-coupled} spins in a DQD. We first introduce the model Hamiltonian of the system and discuss the operation of quantum gates. Then, we present the dephasing of \added{an $S-T_0$} qubit or a two-qubit gate, the methods of suppressing dephasing, and a conclusion in the end.

\section{Model Hamiltonian}


{
Let's consider a gate-defined DQD, where the electric field from the interface and metallic gates confines electrons [see Figure \ref{fig1}(a)]. A two-site Hubbard model can be used to describe the system \cite{hubbard_electron_1963, ashcroft1976},
\beqa
H_0&=&-\sum_{i,\sigma} t_{i,i+1}(\hat{c}_{i,\sigma}^\dag\hat{c}_{i+1,\sigma} + \hat{c}_{i+1,\sigma}^\dag\hat{c}_{i,\sigma}) \nonumber\\
&& +\sum_{i,\sigma} \mu_i \hat{n}_{i,\sigma}  + U\sum_i \hat{n}_{i,\uparrow}\hat{n}_{i,\downarrow},
\eeqa
where ${\hat {c}}_{i,\sigma }^{\dagger }$ (or ${\hat {c}}_{i,\sigma }^{\dagger }$) is the creation (or annihilation) operator for an electron with spin $\sigma$ on the $i$-th site ($i=1$, $2$ for the left or right dot), ${\hat {n}_{i,\sigma }={\hat {c}}_{i,\sigma }^{\dagger }{\hat {c}}_{i,\sigma }}$ is the spin-density operator for spin $\sigma$ on the $i$-th site, $t_{i,j}$ is the tunneling between the neighboring sites, $\mu_i$ is the electron chemical potential on the $i$-th site, and $U$ is the Hubbard energy due to the on-site Coulomb repulsion. 
{Note although the electron system is taken as an example, the model also applies to a hole system.}

{
The device layout and gate potentials can modify the terms in the Hamiltonian. The tunneling is electrically tunable via the gate between the dots [see Figure \ref{fig1}(a)]. The electric chemical potential of each QD can be tuned via the gate on top of each QD. The Hubbard energy depends on the size of the QDs.}
{In the model, we neglect the excited orbital states by assuming that they are high in energy than the spin splittings and do not modify the qualitative behavior of the spin dynamics.} 
{For example, in silicon, there are six valley states of electrons due to the six-fold degeneracy of the conduction band \cite{yang2013, tahan_relaxation_2014, hao_electron_2014, huang_spin_2014, gamble_valley_2016, boross_control_2016, huang_electrically_2017, hwang_impact_2017, zimmerman_valley_2017, schoenfield_coherent_2017, 
mi_high-resolution_2017, salfi_valley_2018, corna_electrically_2018, penthorn_two-axis_2019, borjans_single-spin_2019, zhang_giant_2020, zhang_controlling_2021}. For silicon QDs, the valley degeneracy is lifted by the electric field at the interface. 
The lower two valley states can induce spin relaxation via SOC. However, its magnitude is much less than the spin pure dephasing when the spin and valley splittings are off-resonance} 
\cite{yang2013, tahan_relaxation_2014, hao_electron_2014, schoenfield_coherent_2017, penthorn_two-axis_2019}. We only consider the lowest valley state by assuming that the electron Zeeman splitting and thermal energy are well below the valley splitting. 
Finally, the direct exchange interaction from the Coulomb interaction has been neglected since its magnitude is generally much less than that from the Hubbard model \cite{burkard_coupled_1999, kornich_phonon-mediated_2014, bakker_validity_2015}.

\begin{figure}[t]
\centering
\includegraphics[scale=0.98]{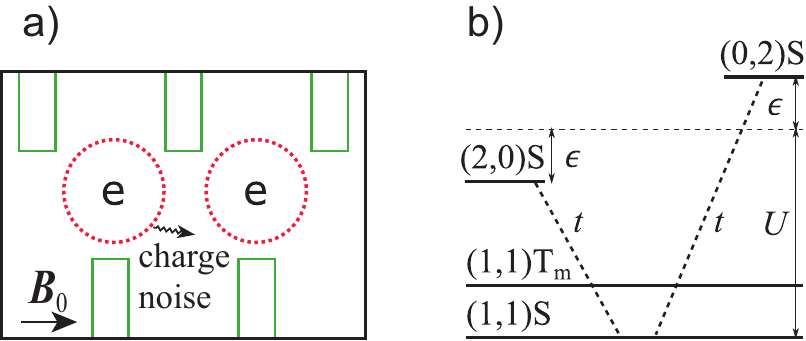}
\caption{Schematic plots of the device and the energy diagram. Panel (a): A schematic of two electron spin qubits in a gate-defined DQD. 
$B_0$ is the applied magnetic field. The ubiquitous charge noise in solid-state devices induces spin decoherence.
Panel (b): Schematic energy diagram of states for two electrons in a DQD, where $\epsilon$ ($t$) is the detuning (tunneling), and $U$ is the Hubbard energy due to Coulomb repulsion. The tunneling couples the singlet states, which shifts the energy of the $\ket{(1,1)S}$ and gives rise to an exchange interaction.
}\label{fig1}
\end{figure}




For two electrons in a DQD, the system Hamiltonian can be written in basis states $\ket{(N_l,N_r)X}$, where $N_l$ ($N_r$ is the number of electrons in the left (right) dot, and $X$ is the spin state of two electrons ($S$ for the spin-singlet and $T_+$, $T_0$, or $T_-$ for the spin-triplet).
In the $(1,1)$ charge regime, $\ket{(1,1)X}$ are the lowest. The tunneling $t_0=t_{1,2}$ between the two dots couples $\ket{(1,1)S}$ to the double occupation states $\ket{(2,0)S}$ and $\ket{(0,2)S}$ [see Figure \ref{fig1}(b)].
Note that according to the Pauli exclusion principle, the spin-triplet states of double occupations are not allowed unless one of the electron occupies an excited orbital state [orbital splitting is assumed big].
%
Then, the system Hamiltonian in the basis $\ket{(1,1)S}$, $\ket{(2,0)S}$, and $\ket{(0,2)S}$ is
\begin{eqnarray}
H_0&=&\left[\begin{array}{cccccc}
0 & \sqrt{2}t_0  & \sqrt{2}t_0 \\
\sqrt{2}t_0 & {U}-\epsilon_0 & 0 \\
\sqrt{2}t_0 & 0 & {U}+\epsilon_0
\end{array}\right], \label{H0}
\end{eqnarray}
where $\epsilon_0=\mu_2-\mu_1$ and $t_0$ are the detuning and tunneling between the two dots. 
Since the tunneling $t_0$ couples $\ket{(1,1)S}$ to the double occupation states $\ket{(2,0)S}$ and $\ket{(0,2)S}$, the state $\ket{(1,1)S}$ is mixed with the double occupation states and becomes $\ket{(1,1)S^\prime}$, whose energy is shifted.
Whereas, there is no such mixing and energy shift for $\ket{(1,1)T_m}$ due to the absence of the triplet states of double occupation as required by the fermion statistics. The energy difference between the singlet and the triplet states results in an exchange interaction.
In the limit $t_0 \ll U \pm \epsilon_0$, the effective Hamiltonian is 
\beq
H_{ex}=\frac{J}{4} (\boldsymbol\sigma_1\cdot\boldsymbol\sigma_2-I),
\eeq
where $J= 2t^2/(U-\epsilon_0) + 2t^2/(U+\epsilon_0)$ is the strength of exchange interaction, and $I$ is the identity operator.
The exchange interaction is purely quantum mechanical effects as required by the antisymmetric fermion-statistics of electrons.
Note the form of exchange-interaction is general. For example, it applies to the exchange interaction arising from the direct Coulomb interaction.
Besides the exchange interaction, there is also Zeeman interaction when a magnetic field is applied. The Zeeman Hamiltonian is $H_{Z}=\sum_i E_{Z,i}\sigma_{z,i}/2$, where $E_{Z,i}$ is the Zeeman splitting of an electron on the $i$-th site. [The effect of Zeeman splitting may also be included in Hubbard model by using a spin-dependent chemical potential, $\mu_{i,\sigma}=\mu_i + E_{Z,i}\sigma_{z,i}/2$]. Note that the Zeeman splittings can be different in dots due to inhomogeneous magnetic field, g-factor difference, or nuclear spins \cite{petta2005, veldhorst2015, kalra_robust_2014}. 
Thus, in the basis $\ket{(1,1)T_{+}}$, $\ket{(1,1)T_0}$, $\ket{(1,1)S}$, $\ket{(1,1)T_{-}}$, the two-spin effective Hamiltonian is
\begin{eqnarray}
H_0&=&\left[\begin{array}{cccccc}
\overline{E}_{Z} & 0 & 0 & 0 & \\
0 & 0 & \frac{\delta E_Z}{2} & 0 \\ 
0 & \frac{\delta E_Z}{2} & -J & 0 \\
0 & 0 & 0 & -\overline{E}_{Z}
\end{array}\right], \label{H0eff}
\end{eqnarray}
where $\delta E_Z=(E_{Z,1}-E_{Z,2})$ (or $\overline{E}_Z=(E_{Z,1}+E_{Z,2})/2$) is the difference (or the average) of the Zeeman splittings.
The Hamiltonian will form the basis for the following study of spin manipulation and decoherence.
%


In the following, we discuss the most widely used quantum gates based on the exchange interaction and how charge noise can affect the spins through the exchange interaction.

%
%


\section{Quantum gates based on the exchange interaction}

%
%
%
%

It is an essential but challenging task to realize two-qubit gate operations for a quantum computer.
For spin qubits, two-qubit gates are operated based on exchange interaction, strong coupling of spin and photon, superexchange coupling, or capacitive coupling \cite{veldhorst2015, zajac_resonantly_2018, watson_programmable_2018, hendrickx_fast_2020, mi_coherent_2018, samkharadze_strong_2018, baart_coherent_2017, nichol_high_fidelity_2017}.
The two-qubit gates demonstrated in recent experiments, including the SWAP gate, the CPHASE gate, and the CROT gate, are mediated by the exchange interaction between electrons in the nearby QDs in silicon \cite{veldhorst2015, zajac_resonantly_2018, watson_programmable_2018, huang_fidelity_2019, xue_benchmarking_2019, sigillito_coherent_2019, he_two-qubit_2019, hendrickx_fast_2020}. 
Moreover, there are different qubit realizations and encoding schemes based on the exchange interaction. 
Next, we discuss how the two-qubit gates and qubit encodings are realized based on the exchange interaction. 




\subsection{Two-qubit gates based on exchange interaction}

Here, we discuss the realization of two-qubit gates, including the $\sqrt{\text{SWAP}}$ gate, the CPHASE gate, and the CROT gate.

\textbf{$\sqrt{\text{SWAP}}$ gate---}
$\sqrt\text{SWAP}$ gate is one of the most important two-qubit entangling gates that can be realized using tunable exchange interaction. Based on the $\sqrt\text{SWAP}$ gate and single-qubit rotations, it is possible to construct universal quantum gates for quantum computing \cite{loss1998}. SWAP gate can also transfer quantum information among qubits for measurement or realizing remote quantum gate \cite{qiao_conditional_2020}. 
%
For the operation of two-qubit gate, the computational basis states of two qubits can be defined as $\ket{\uparrow\uparrow}=\ket{T_+}$, $\ket{{\uparrow\downarrow}^\prime}=\ket{T_0}+\ket{{S}^\prime}$, $\ket{{\downarrow\uparrow}^\prime}=\ket{T_0}-\ket{{S}^\prime}$, and $\ket{\downarrow\downarrow}=\ket{T_+}$ [notation (1,1) is omitted for simplicity]. In this case, an ideal $\sqrt{\text{SWAP}}$ gate is rotation about the $J$ axis of an angle $\pi$ \cite{nielsen2010, kalra_robust_2014}. 
A scheme of realizing a $\sqrt{\text{SWAP}}$ gate is to have a finite magnetic field gradient between the two dots \cite{kalra_robust_2014}. 
Then, by tuning the exchange interaction, one can realize a high-fidelity $\sqrt{\text{SWAP}}$  gate. 
It has been shown that the minimum gate fidelity can be as high as 99\%, when the exchange is tunable from $0.1 \delta E_Z$ to $10\delta E_Z$ when decoherence negligibly slow \cite{kalra_robust_2014}. {Note the exchange coupling is tunable by orders of magnitude by changing either the detuning or the tunneling \cite{petta2005, veldhorst2015, bertrand_quantum_2015, reed_reduced_2016, martins_noise_2016,  watson_programmable_2018, zajac_resonantly_2018}.}
High-fidelity SWAP gates have been realized recently in QD systems \cite{he_two-qubit_2019, sigillito_coherent_2019, qiao_conditional_2020}. 
%

%

\textbf{CPHASE gate---}
A CPHASE gate is another entangling gate used to construct universal quantum gates up to single-qubit rotations.
An ideal CPHASE gate is defined as a z-rotation by a phase angle on target-qubit when control-qubit is $\ket{1}$ (i.e. spin-up). 
A C-PHASE gate can also be realized by tuning the exchange interaction \cite{meunier_efficient_2011, veldhorst2015}.

Suppose $J<\delta E_Z\ll \overline{E}_Z$ , then, through a rotation in the subspace $\ket{\uparrow\downarrow^\prime}$ and $\ket{\downarrow\uparrow^\prime}$, the Hamiltonian can be diagonalized as
\begin{eqnarray}
H_0=\left[\begin{array}{cccccc}
\overline{E}_{Z} & 0 & 0 & 0 & \\
0 & \frac{\delta E_Z^\prime}{2}-\frac{J}{2} & 0 & 0 \\ 
0 & 0 & -\frac{\delta E_Z^\prime}{2}-\frac{J}{2} & 0 \\
0 & 0 & 0 & -\overline{E}_{Z}
\end{array}\right], \label{H0eff}
\end{eqnarray}
where $\delta E_Z^\prime=\sqrt{\delta E_Z^2 + J^2}$. Correspondingly, the spin eigenstates are $\ket{\uparrow\uparrow}$, $\ket{\downarrow\downarrow}$, $\ket{{{\uparrow\downarrow}^{\prime\prime}}}=\cos\frac{\theta_{ST}}{2}\ket{\uparrow\downarrow^\prime} -\sin\frac{\theta_{ST}}{2} \ket{\downarrow\uparrow^\prime}$, and $\ket{{{\downarrow\uparrow}^{\prime\prime}}}=\sin\frac{\theta_{ST}}{2}\ket{\uparrow\downarrow^\prime} +\cos\frac{\theta_{ST}}{2} \ket{\downarrow\uparrow^\prime}$, where $\theta_{ST}=\tan^{-1}(J/\delta E_Z)$ is the {angle due to the exchange coupling induced state mixing}.
In the limit of $J \ll \delta E_Z$, the control-qubit and target-qubit are well defined.
Suppose the electron spin in the left dot (right dot) is the control-qubit (target-qubit), then the control-qubit effectively selects a subspace of the system (see Figure \ref{Fig_scheme}).
If the control-qubit is spin-up (or spin-down), then, it selects a subspace of $\ket{\uparrow\uparrow}$ and $\ket{\uparrow\downarrow^{\prime\prime}}$ with an energy splitting $E_{\uparrow\updownarrow} = \overline{E}_Z -\delta E_Z^\prime/2 + J/2$ (or subspace of $\ket{\downarrow\uparrow^{\prime\prime}}$ and $\ket{\downarrow\downarrow}$ with an energy splitting $E_{\downarrow\updownarrow}=\overline{E}_Z -\delta E_Z^\prime/2 -J/2$).

A CPHASE gate can be realized by performing a Ramsey-like sequence on the two-spin system. {Consider the two-spin is initialized to a state $\ket{\downarrow +}=\frac{1}{\sqrt{2}}(\ket{\downarrow\downarrow}+\ket{\downarrow\uparrow^{\prime\prime}})$ when the exchange $J$ is off [i.e. one first initializes the two spins to the ground state $\ket{\downarrow\downarrow}$, then, a microwave pulse on a metallic gate can induce a $\pi/2$ $x$-rotation on the target-qubit, which prepares the two-spin in a state $\ket{\downarrow +}$]. Suppose the exchange interaction $J$ is then turned on for $t_{on}$, then, the target-qubit undergoes a precession of frequency $(\Delta-J/2)/\hbar$, where $\Delta/\hbar=E_{Z,2}/\hbar- \omega_0$ is the detuning between the frequency of the target qubit and the frequency $\omega_0$ of the driving field. However, if the initial state is $\ket{\uparrow+}=\frac{1}{\sqrt{2}}(\ket{\uparrow\downarrow^{\prime\prime}}+\ket{\uparrow\uparrow})$ with the control qubit being spin-up, then, the precession frequency after turning on exchange is $(\Delta+J/2)/\hbar$.
The precession frequency of the target qubit is different when the control-qubit state is different}, and the resulting phase difference $\phi=J t_{on}/\hbar$ leads to a CPHASE gate [up to a single-qubit rotation].
%
Thus, the speed of the CPHASE gate is proportional to $J$, and a CZ gate can be realized within the time of $t_{on}=\pi\hbar/J$. CPHASE gates have been realized recently in QD systems \cite{veldhorst2015}.

\begin{figure}[t]
\centering
\includegraphics[scale=1.5]{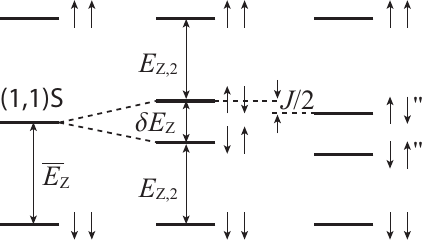}
\caption{Schematic energy diagram of two-electron spin states when $\delta E_Z=J=0$, $\delta E_Z \ne J=0$, or $\delta E_Z >J\ne0$. The exchange interaction $J$ lowers the unpolarized two-spin states essential for the CPHASE and CROT gates.
}\label{Fig_scheme}
\end{figure}

%
%


\textbf{CROT gate---}
For universal quantum computing, an alternative entangling gate is the CROT gate. The CROT gate is an $x$-rotation of the target-qubit when the control-qubit is $\ket{1}$ (i.e., spin-up).
Through a combination of the single-spin rotation and exchange interaction, a CROT gate can be realized \cite{kalra_robust_2014, zajac_resonantly_2018}. {In the presence of the electron spin resonance (or \added{EDSR}) that provides a single-spin rotation \cite{koppens_driven_2006, rashba_orbital_2003, flindt_spin-orbit_2006, golovach2006, nowack_coherent_2007, rashba_theory_2008, tokura_coherent_2006, pioro-ladriere_micromagnets_2007, pioro-ladriere_electrically_2008, kawakami_electrical_2014, wu_two-axis_2014, yoneda2018}}, the system Hamiltonian becomes
\begin{eqnarray}
H=\left[\begin{array}{cccccc}
\overline{E}_{Z} & c_a\Omega_s(t) & c_b\Omega_s(t) & 0 & \\
c_a\Omega_s(t) & \frac{\delta E_Z^\prime-J}{2} & 0 & c_a\Omega_s(t) \\ 
c_b\Omega_s(t) & 0 & \frac{-\delta E_Z^\prime-J}{2} & c_b\Omega_s(t) \\
0 & c_a\Omega_s(t) & c_b\Omega_s(t) & -\overline{E}_{Z}
\end{array}\right], \label{H0eff}
\end{eqnarray}
where $\Omega_s(t)=\Omega_Re^{i\omega_0 t}$ is due to the single-qubit driving, and the parameters $c_a=\cos\theta_{ST}-\sin\theta_{ST}$ and $c_b=\cos\theta_{ST}+\sin\theta_{ST}$ is due to the mixing of states $\ket{\uparrow\downarrow^\prime}$ and $\ket{\downarrow\uparrow^\prime}$.
\added{Suppose the condition $J\ll \delta E_Z$ is satisfied, then, $\theta_{ST}\approx 0$ and $c_a\approx c_b\approx 1$.}

As discussed in the case of the CPHASE gate, the exchange interaction leads to the different splitting of the target-qubit for different states of the control-qubit. If a microwave of frequency \added{$(\overline{E}_Z-\delta E_Z^\prime/2 + J/2)/\hbar$} is applied, then, the target-qubit is driven resonantly only when the control-qubit is spin-up.
In this case, the target-qubit undergoes a single-qubit rotation, $e^{-i\Omega_R t \sigma_{x,2}/(2\hbar)}$. However, if the control-qubit is spin-down, the driving field is off-resonance with the target-qubit, the target-qubit undergoes a different rotation, $e^{-it (\Omega_R \sigma_{x,2} - J\sigma_{z,2})/(2\hbar)}$. Therefore, a CROT gate can be realized when $\Omega_R\ll J$ and $Jt=2\pi\hbar n$ with $n$ an integer number. In particular, a controlled-not (CNOT) gate is realized in time $t=\pi\hbar/\Omega_R$. Note if $\Omega_R/J$ is finite, there will be an unwanted rotation of the target-qubit when the control-qubit is spin-down.
In this case, echo pulses may help improve the gate fidelity. CROT gates have been realized recently in QD systems \cite{zajac_resonantly_2018}.





\subsection{Donor-based QDs and $S-T_0$ qubit encoding}

Besides the gated QDs formed from a 2DEG, donor-based QD can also be fabricated using either ion implantation \cite{shinada_enhancing_2005, sellier_transport_2006, lansbergen_gate-induced_2008, morello_single-shot_2010, johnson_drain_2010, roche_detection_2012, prati_anderson-mott_2012, pla_high-fidelity_2013, dehollain_bells_2016, tosi_silicon_2017, harvey-collard_coherent_2017, harvey-collard_high-fidelity_2018, jock_silicon_2018, harvey-collard_spin-orbit_2019, asaad_coherent_2020, morello_donor_2020, madzik_conditional_2021} or STM lithography using a scanning probe microscope \cite{ruess_toward_2004, simmons_scanning_2005, fuechsle_spectroscopy_2010, fuechsle_single-atom_2012, weber_ohms_2012, watson_transport_2014, broome_two-electron_2018, koch_spin_2019, he_two-qubit_2019}. Moreover, the spin states in QDs can be used to encode the so-called logical \added{$S-T_0$} qubits \cite{petta2005, taylor_fault-tolerant_2005, maune_coherent_2012, shulman_demonstration_2012, wu_two-axis_2014, nichol_high_fidelity_2017, kandel_coherent_2019, qiao_conditional_2020, qiao_coherent_2020, qiao_floquetenhanced_2021}. The analysis of spin manipulation and dephasing can be applied to donor-based spin qubits or a logical $S-T_0$ qubit.
Here, we briefly discuss the two-electron spin qubits in donors and the encoding of logical spin qubits.

\textbf{Donor-based QDs---}
An alternative physical realization of spin qubits is based on the donors in silicon \cite{morello_donor_2020, he_two-qubit_2019, wellard_electron_2003, calderon_quantum_2006, rahman_high_2007}. The positive charge of the donor ion provides confinement for an electron. The electron spin in a donor atom has a long coherence time due to the strong confinement \cite{muhonen2014}. The donor nuclear spin also couples to the electron spin through the hyperfine interaction \cite{zwanenburg_silicon_2013}. Electron spin splitting varies in two nearby donors due to the hyperfine interaction and the initialized nuclear spins. Tunable exchange interaction is achieved through the energy detuning between the two donors \cite{kalra_robust_2014}. Thus, for spin qubits in donor atoms, the Hamiltonian and two-qubit gates mentioned above can be realized similarly.



\textbf{$S-T_0$ qubit encoding---}
Besides the two-qubit gate based on the exchange interaction, the so-called logical spin qubit can also be defined, where a qubit is encoded in the two states of multiple electron (or hole) spins. The most well-known logical qubit is the $S-T_0$ qubit, where the qubit is defined by the superpositions of state $\ket{(1,1)S}$ and $\ket{(1,1)T_0}$ \cite{petta2005, taylor_fault-tolerant_2005}. In this case, the inhomogeneous Zeeman splitting and the exchange interaction define the two independent rotation axis. The exchange interaction is used as a single-qubit rotation for the $S-T_0$ qubit. In the case of $S-T_0$ qubits, a two-qubit gate can be realized by using the capacitive coupling or the exchange interaction \cite{shulman_demonstration_2012, nichol_high-fidelity_2017, qiao_floquetenhanced_2021}. In addition to the $S-T_0$ qubit, different logical spin qubits can also be defined in QDs, such as exchange-only qubit, resonant exchange qubit, hybrid spin qubit, etc. \cite{divincenzo_universal_2000, laird_coherent_2010, medford_selfconsistent_2013, medford_quantumdotbased_2013, malinowski_symmetric_2017, shi_fast_2012, shi_fast_2014, kim_high-fidelity_2015, kim_microwave-driven_2015, cao_tunable_2016}.

\subsection{Summary of exchange-based quantum gates}

In summary, based on the exchange interaction, various two-qubit gates, including the SWAP, the CPHASE, and the CROT can be realized for two-electron spins in either a gate-defined or donor-based DQD. The same physical system can also be used as a high-quality logical spin qubit, such as the $S-T_0$ qubit.

\section{Decoherence mechanism of a two-spin $S-T_0$ qubit in QDs}

Despite the importance of the exchange interaction for the realization of quantum gates of spin qubits, the exchange interaction also introduces decoherence channels in the presence of noise. Since spin pure dephasing is generally much faster than spin relaxation, we focus on the pure spin dephasing of two-electron spins in a DQD. Low-frequency charge noise causes the fluctuation of the exchange interaction and leads to severe spin pure dephasing \cite{hu2006}. The dephasing via the exchange interaction has been studied intensively in the literature \cite{hu2006, laird_effect_2006, bermeister_charge_2014, tahan_relaxation_2014, huang_electron_2014, huang_spin_2014, kha_micromagnets_2015, huang_spin_2018, hollmann_large_2020, struck_low-frequency_2020} and schemes have been proposed to protect qubits against charge noise \cite{salfi_quantum_2016, shim_charge-noise-insensitive_2016, nichol_high-fidelity_2017, zhang_randomized_2017, throckmorton_fast_2017, friesen_decoherence-free_2017, yang_suppression_2017, shim_barrier_2018, russ_quadrupolar_2018, benito_electric-field_2019}. In the following, we report progress on the study of spin pure dephasing due to charge noise in a DQD. We first introduce the electric noises in the system. Then, we discuss the dephasing of an $S-T_0$ qubit before considering the dephasing in a two-qubit gate.

\subsection{Noise in the system}

%

A quantum state can be coherent superpositions of states. However, uncontrollable environmental noise destroys the quantum coherence and limits the quantum gate fidelity.

For a spin qubit in solid-state QDs, there could be magnetic or electrical noises. Magnetic noise, such as nuclear spin fluctuations, is known to be detrimental to spin qubits in QDs \cite{hanson_spins_2007}. The nuclear noise can be mostly removed in isotopically enriched $^{28}$Si \cite{tyryshkin_electron_2003, tyryshkin_electron_2012}. On the other hand, the electrical noise is also ubiquitous in solid states. One of the most notorious source of decoherence arises from the low frequency $1/f$ charge noise \cite{dutta_lowfrequency_1981, weissman_frac1f_1988, jung_background_2004, astafiev_quantum_2004, zimmerman_why_2008, dial_charge_2013,zimmerman_charge_2014, freeman_comparison_2016, shamim_ultralow-noise_2016, connors_low-frequency_2019, kranz_exploiting_2020}.
Electric fields are not interacting with spin qubits directly. However, they affect spin qubits indirectly via the exchange interaction or the \added{SOC}
 \cite{hu2006, huang_electron_2014, huang_spin_2014, kha_micromagnets_2015, qi_effects_2017, huang_spin_2018, yang_high-fidelity_2019, borjans_single-spin_2019, hollmann_large_2020, struck_low-frequency_2020, huang_fast_2020, huang_impact_2020, zhang_giant_2020, zhang_controlling_2021}. In the following, we focus on the spin decoherence due to charge noise through the exchange interaction, which is the most important in a two-spin system in a DQD. 


For QD devices, charge noise can be measured by the Coulomb blockade spectrum of a single electron transistor, where the electron chemical potential $\delta \mu$ is shifted by the electric noise \cite{zimmerman_why_2008}. The spectral density
$S_{1/f}(\omega)=\int_{-\infty}^{\infty}\langle \delta \mu (0) \delta \mu(\tau) \rangle \cos(\omega\tau)d\tau$ of the $1/f$ charge noise is
\beq
S_{1/f}(\omega) = A/\omega,
\eeq
where $A$ is the amplitude of charge noise \cite{dutta_lowfrequency_1981, weissman_frac1f_1988}. Although $S_{1/f}(\omega)$ usually exhibits $1/\omega$ dependence, the exponent of $\omega$ can be vary from 0 to 2 for different devices. The energy fluctuations $\sqrt{A}$ goes from 0.1 $\mu$eV to 10 $\mu$eV depending on material and experimental details \cite{dutta_lowfrequency_1981, weissman_frac1f_1988, zimmerman_charge_2014, freeman_comparison_2016}, {and it also increases with the temperature of the environment\cite{connors_low-frequency_2019}.}

{Besides the charge noise, there are also phonon and Johnson noises \cite{langsjoen_qubit_2012, poudel_relaxation_2013, huang_electron_2014, tahan2014}. 
Phonon or Johnson noise induces spin relaxation, which is temperature-dependent and usually much less than the overall spin pure dephasing \cite{khaetskii2001, golovach2004, yang2013, tahan_relaxation_2014, huang_electron_2014, huang_spin_2014}.
At low temperatures ($k_BT < \overline{E}_Z$), the spin pure dephasing due to phonon or Johnson noise is negligible due to the small spectral density at low frequencies \cite{golovach2004, huang_electron_2014}. At an elevated temperature, there could be sizable spin pure dephasing due to multi-phonon emission \cite{kornich2014}. The spin pure dephasing at low temperatures is mostly from charge noise, which is the main focus below.}



%

\subsection{Dephasing of an $S-T_0$ qubit from electric noise}

Here, we discuss the effect of charge noise on an $S-T_0$ qubit, which mainly follows Ref. \cite{hu2006}.
Suppose charge noise causes a fluctuating chemical potential $\delta \mu$ in each dot,
then, the detuning of a DQD is shifted by $\hat{n}_\epsilon\approx \delta \mu$ due to the coupling to the charge noise.
[Charge noise also induces tunneling fluctuations, which leads to spin dephasing and will be discussed later on.]
As the detuning $\epsilon$ varies, the exchange interaction is modified as $J=J_0(\epsilon) + (\partial J/\partial \epsilon) \hat{n}_\epsilon$, where $\partial J/\partial \epsilon$ measures the sensitivity of the exchange interaction to the detuning. Therefore, the system Hamiltonian including noise is
\begin{eqnarray}
H^\prime&=&\left[\begin{array}{ccccc}
0 & \frac{\delta E_Z}{2} \\ 
\frac{\delta E_Z}{2} & -J_0-\hat{n}_{}^\prime \\
\end{array}\right],
\end{eqnarray}
where $J_0\approx \frac{2t_0^2}{U-\epsilon_0} + \frac{2t_0^2}{U+\epsilon_0}$ is the exchange interaction, and the noise is $\hat{n}_{}^\prime= (\partial{J}/\partial \epsilon_0) \hat{n}_\epsilon = [\frac{2t_0^2}{(U-\epsilon_0)^2} - \frac{2t_0^2}{(U+\epsilon_0)^2}]n_{\epsilon}$.
%

Suppose the random detuning noise $\hat{n}_\epsilon$ has Gaussian distribution when the noise is slowly varying, the resulting pure dephasing between the states of an $S-T_0$ qubit can be evaluated \cite{duan_reducing_1998, taylor_dephasing_2006},
$\overline{\exp(i\delta_\phi)}=\exp[-\phi(t)]$, where
\beqa
\phi(t)&=& \int_{\omega_0}^{\infty} d\omega J_{zz}(\omega) [2\sin(\omega t/2)/\omega]^2, \label{phit}
\eeqa
$J_{zz}(\omega)$ the is spectral density of the relative noise $\hat{h}_z=\hat{n}^\prime/2$ between the states of interest, $J_{zz}(\omega)= \frac{2}{\hbar^2}\int_{-\infty}^{\infty}\langle \hat{h}_{z}(0)\hat{h}_{z}(\tau) \rangle \cos(\omega\tau)d\tau$,
and $\omega_0$ is the cutoff frequency set by the measurement time.
%
By evaluating the amplitude of the noise and its effect on exchange interaction, the pure dephasing $1/T_\varphi$ between the states $\ket{(1,1)S}$ and $\ket{(1,1)T_0}$ can be evaluated, which is defined as $\phi(t=T_\varphi)=1$.
Note that the expression is generally applicable for the dephasing between arbitrary two states. Suppose there are two states $\ket{\alpha}$ and $\ket{\beta}$ of interest, then, $J_{zz}(\omega)$ in the Eq. (\ref{phit}) is replaced by the power spectral density $J_{zz}^{\alpha\beta}(\omega)$ of the relative noise $\hat{h}_z^{\alpha\beta}=(\hat{n}_{\alpha\alpha}-\hat{n}_{\beta\beta})/2$, where $\hat{n}_{\alpha\alpha}$ (or $\hat{n}_{\beta\beta}$) is the noise on the state $\ket{\alpha}$ (or $\ket{\beta}$).

\begin{figure}[t]
\centering
\includegraphics[scale=0.75]{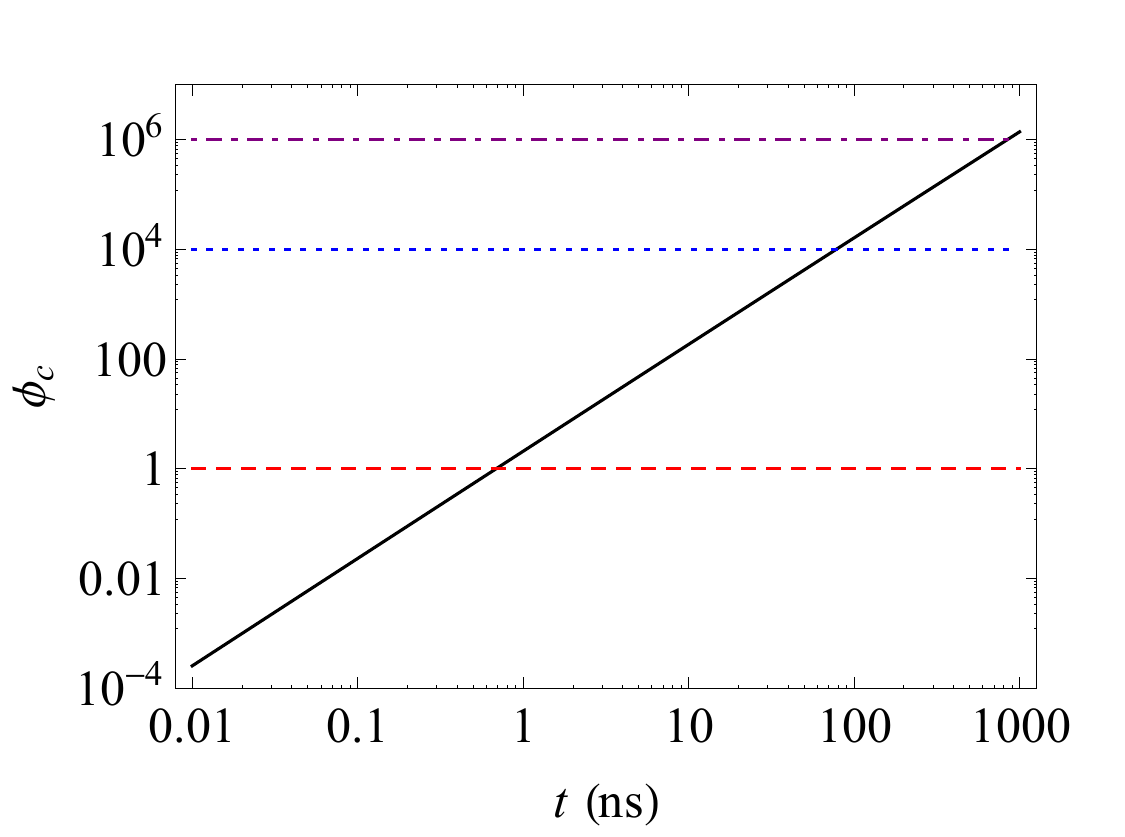}
\caption{Phase diffusion $\phi_c$ as a function of time for a charge qubit when $t_0\ll |\epsilon_0|$ \added{(Reproduced with permission. \cite{hu2006} 2021, APS)}. The cutoff frequency is $\omega_0=1$ Hz. The dephasing time is around 1 ns for a charge qubit [corresponds to $\phi_c=1$]. Dephasing of an $S-T_0$ qubit can be obtained accordingly, since $\phi_{ST_0}=(\partial J/\partial \epsilon)^2\phi_c$. If $dJ/dV=0.01$ (or 0.001), the dephasing time of an $S-T_0$ qubit is 0.1 $\mu$s (or 1 $\mu$s) [see the horizontal lines where $\Delta \phi_c \sim 10^4$ $(10^6)$].
}\label{Fig_phi_c}
\end{figure}

Before evaluating the dephasing of an $S-T_0$ qubit, we consider the dephasing of a charge qubit in a DQD. The Hamiltonian of a charge qubit is $H =\epsilon_0 \sigma_z + t_0 \sigma_x$, where $\epsilon_0$ (or $t_0$) is the detuning (or tunneling) between the two dots.
%
The phase diffusion $\phi_c$ can be evaluated for a far-detuned charge qubit [i.e. when $t_0\ll \epsilon_0$],
\beqa
\phi_c(t)&=& \int_{\omega_0}^{\infty} d\omega S_{1/f}(\omega) [2\sin(\omega t/2)/\omega]^2, \label{phi_c}
\eeqa
where $S_{1/f}(\omega)$ is the spectral density introduced above. For the $1/f$ charge noise, analytical results suggest that $\phi_c \propto t^2 \ln t$, which has a nearly quadratic dependence on time. Numerical evaluation indicates that the dephasing of a charge qubit is as fast as 10 ns, when the charge relaxation time is 10 ns \cite{hu2006} [see Figure \ref{Fig_phi_c}].

We now look at the dephasing of a two-spin $S-T_0$ qubit, where the effective Hamiltonian can be written as $H=J_0\sigma_z$ ($\delta E_Z=0$ is assumed). The phase diffusion of the $S-T_0$ qubit is proportional to that of a charge qubit, 
\beq
 \phi_{ST0}=(\partial J/\partial \epsilon)^2\phi_c. \label{phi_s}
\eeq
Note that a SWAP gate is operated in the limit $\delta E_Z \ll J_0$. In this limit, the dephasing of an $S-T_0$ qubit also corresponds to that of a SWAP gate.
The dephasing of an $S-T_0$ qubit depends on the parameter $\partial J/\partial \epsilon$, which varies widely (see Figure \ref{Fig_dEde} and Ref. \cite{petta2005}). If $dJ/d\epsilon=1$ (when $t\approx U-\epsilon$), the dephasing is as short as {the dephasing of} a charge qubit. However, if $dJ/d\epsilon=0.01$ (or 0.001), the dephasing would be 100 (or 1000) times slower, leading to $T_\varphi=0.1$ $\mu$s (or 1 $\mu$s) \cite{hu2006} [see the blue dashed and the purple dash-dotted lines in Figure \ref{Fig_phi_c}]. 
Thus, one needs a small value of $dJ/d\epsilon$ to suppress the dephasing of an $S-T_0$ qubit from the detuning fluctuations. Experimental studies have shown indeed charge noise has a significant contribution to the dephasing of an $S-T_0$ qubit \cite{laird_effect_2006}.

\subsection{Suppressing the dephasing of an $S-T_0$ qubit}

Recent studies reveal that an an effective way to suppress the spin dephasing is the so called symmetric operation \cite{bertrand_quantum_2015, reed_reduced_2016, martins_noise_2016}. 
Another method is to use inhomogeneous Zeeman splitting to suppress the spin pure dephasing \cite{nichol_high-fidelity_2017}. We will discuss them in the following.

\subsubsection{Reduced sensitivity though symmetric operation}

\begin{figure}[]
\centering
\includegraphics[scale=0.75]{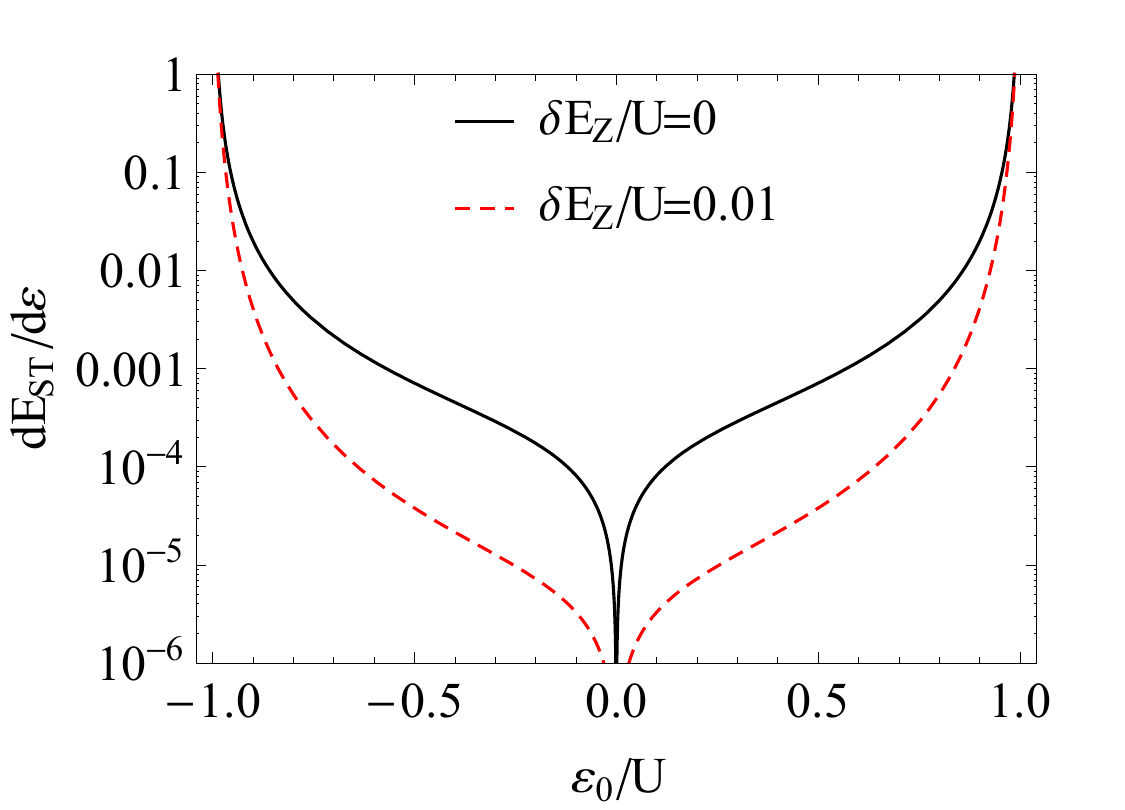}
\caption{Sensitivity $dE_{ST}/d\epsilon$ of the energy of an $S-T_0$ qubit to the detuning as a function of the detuning $\epsilon$ when $\delta E_Z=0$ (black solid line) or $\delta E_Z=0.01U$ (red dashed line) [$t_0=0.01U$]. When $\delta E_Z=0$, $dE_{ST}/d\epsilon=dJ/d\epsilon$ is minimized at the symmetric point $\epsilon_0 = 0$. As $\delta E_Z$ increases to $0.01U$, the sensitivity $dE_{ST}/d\epsilon$ is reduced in wide range of detuning.
}\label{Fig_dEde}
\end{figure}

%

To suppress the dephasing of an $S-T_0$ qubit due to detuning noise, one can operate the system at the symmetric operation point, where the detuning of the DQD is zero.
Figure \ref{Fig_dEde} shows the sensitivity $dE/d\epsilon$ as a function of detuning $\epsilon$, in which $dE/d\epsilon$ is minimized for both $\delta E_Z=0$ and $\delta E_Z=0.01 U$ at the symmetric operation point. In particular, when $\delta E_Z=0$, where the energy splitting of an $S-T_0$ qubit is solely determined by the exchange interaction $J$, the parameter $dE/d\epsilon=dJ/d\epsilon$ is minimized at zero detunings (i.e., symmetric operation point). Thus, according to Eq. \ref{phi_s}, the dephasing due to detuning fluctuation induced by charge noise is minimized at the symmetric operation point.
As shown in several recent experiments, the coherence time $T_2^*$ of an $S-T_0$ qubit can be improved by more than a factor of 5 due to the reduced sensitivity and the number of exchange oscillations is increased \cite{bertrand_quantum_2015, reed_reduced_2016, martins_noise_2016}.
Note that in the case of symmetric operation, exchange remains highly tunable by adjusting the tunnel coupling \cite{bertrand_quantum_2015, reed_reduced_2016, martins_noise_2016}.
%



\subsubsection{Reduced sensitivity though Zeeman splitting difference}

Alternatively, the difference $\delta E_Z$ of Zeeman splittings can be used to reduce the dephasing of an $S-T_0$ qubit \cite{nichol_high-fidelity_2017}. The Zeeman difference $\delta E_Z$ could be due to the nonuniform magnetic field, Overhauser field, or g-factor in different dots \cite{petta2005, veldhorst2015, zajac_resonantly_2018}.
In the presence of a finite $\delta E_Z$, the energy splitting of an $S-T_0$ qubit is $\delta E_Z^\prime =\sqrt{\delta E_Z^2 + J^2(\epsilon)}$. Thus, the sensitivity $dE/d\epsilon=(dE/dJ)(dJ/d\epsilon)$ decreases as $J/\delta E_Z$ decreases [note $dE/dJ=J/\delta E_Z^\prime$]. Figure \ref{Fig_dEde} shows that, compared with the case of $\delta E_Z=0$, the sensitivity is indeed reduced in a wide range of detuning $\epsilon$ when $\delta E_Z=0.01 U$.
Recent experiments show that, by reducing the exchange interaction $J_0$ relative to $\delta E_Z$, the coherence time $T_2^*$ of an $S-T_0$ qubit is substantially increased \cite{nichol_high-fidelity_2017}.


\section{Decoherence in a two-qubit gate}

In this section, we consider the decoherence in a two-qubit gate. We mainly focus on the dephasing in a two-qubit CPHASE gate due to charge noise mediated by exchange interaction is discussed. It follows mostly one of our previous studies \cite{huang_spin_2018}.

\subsection{Dephasing in a two-qubit CPHASE gate}

For the dephasing in a CPHASE gate, the phases between each pair of the four two-qubit states are relevant. 
[In particular, when the control-qubit is in the eigenstate, the phase difference between the two target-qubit states is important.] 
The exchange interaction induces not only the dephasing between the two unpolarized states, but also the dephasing between the states $\ket{\uparrow\uparrow}$ and $\ket{\uparrow\downarrow}$, or the states $\ket{\downarrow\uparrow}$ and $\ket{\downarrow\downarrow}$. Interestingly, it acts like single-spin dephasing. As discussed above, the Zeeman energy difference would reduce the dephasing between the two unpolarized states. It also leads to the so-called transverse sweet spot (TSS) in the literature for an $S-T_0$ qubit \cite{abadillo-uriel_enhancing_2019}. 
However, the Zeeman energy difference is no longer helpful for a two-qubit gate, especially for the dephasing between a polarized spin state and an unpolarized spin state. 
On the other hand, the symmetric operation point can still reduce the sensitivity to charge noise for a two-qubit gate because it reduces the sensitivity of exchange interaction $J$ to the detuning $\epsilon$ of a DQD.

\begin{figure}[t]
\centering
\includegraphics[scale=0.88]{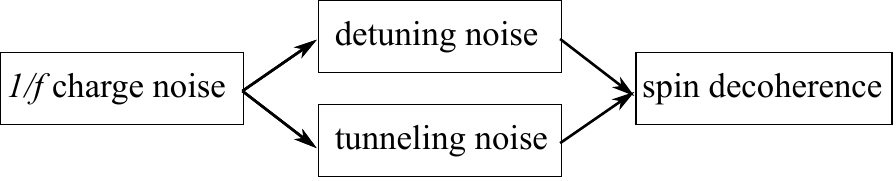}
\caption{Effect of $1/f$ charge noise on spin decoherence in a two-qubit logic gate \cite{huang_spin_2018}.}\label{Fig_noise}
\end{figure}

Besides the detuning noise studied intensively in the literature, there can be fluctuation of tunneling due to charge noise through the variation of the height and width of the tunneling barrier \added{(see Figure \ref{Fig_noise})}. 
Suppose the fluctuation of tunneling barrier is similar in magnitude as the detuning fluctuation, $\delta E_b\approx n_\epsilon$, then, the tunneling noise is estimated as $n_t=(\partial t/\partial E_b)n_\epsilon$, where $\partial t/\partial E_b$ measures the sensitivity of tunneling rate to changes of the height of the tunneling barrier.
The amplitude of tunneling fluctuation is generally weaker than the detuning fluctuation. In the WKB approximation \cite{huang_spin_2018}, 
\beq
\partial t_0/\partial E_b \approx t_0/(2\Delta_b), \label{dtdE}
\eeq
where $\Delta_b$ depends on the height and width of the tunnel barrier between the dots. 
Since $t_0\ll\Delta_b$ is usually satisfied, the tunneling noise is in general much less than the detuning noise. 
It means that for a charge qubit, the tunneling noise is usually weak and neglected. However, as shown below, the tunneling noise contributes significantly to the dephasing of a spin qubit. 

Now, we consider the dephasing in the two-qubit gate due to charge noise. The two-qubit Hamiltonian in the presence of charge noise
in the subspace ($\ket{(1,1)T_0}$, $\ket{(1,1)S^\prime}$) is  
[the polarized spin states $\ket{(1,1)T_\pm}$ are decoupled from the subspace] \cite{huang_spin_2018}
\begin{eqnarray}
H^\prime&=&\left[\begin{array}{ccccc}
0 & \frac{\delta E_Z}{2} \\ 
\frac{\delta E_Z}{2} & -J-\hat{n}_{}^\prime \\
\end{array}\right],
\end{eqnarray}
where the noise $\hat{n}_{}^\prime=(\partial{J}/\partial t_0) \hat{n}_t + (\partial{J}/\partial \epsilon_0) \hat{n}_\epsilon=\sqrt{2}{{\theta}}\hat{n}_{t} - ({{\theta}}^2/4)\hat{n}_{\epsilon}$ and the factor $\theta\approx t_0/(U-\epsilon_0)$ captures the admixture between the $\ket{(1,1)S}$ and $\ket{(2,0)S}$.
Therefore, the effective noise due to the tunneling fluctuation is first order in the small charge admixture, $\partial J/\partial t_{0} \propto \theta$; while the noise due to the fluctuation of the detuning is a second-order effect, $\partial J/\partial \epsilon_0 \propto \theta^2$.
Even if the tunneling noise is smaller than the detuning noise, 
the effect of tunneling noise can still be dominant.

The Hamiltonian $H^\prime$ without noise can be diagonalized, and the eigenstates are denoted as $\ket{\uparrow\downarrow^{\prime\prime}}$ and $\ket{\downarrow\uparrow^{\prime\prime}}$. Then, the effective Hamiltonian becomes
\begin{eqnarray}
H^{\prime\prime}&=&\left[\begin{array}{ccccc}
-\frac{J}{2} + \frac{\Omega_J}{2} +\hat{n}_{\uparrow\downarrow^{\prime\prime},\uparrow\downarrow^{\prime\prime}} & -\hat{n}_{\uparrow\downarrow^{\prime\prime},\downarrow\uparrow^{\prime\prime}} \\ 
-\hat{n}_{\downarrow\uparrow^{\prime\prime},\uparrow\downarrow^{\prime\prime}} & -\frac{J}{2} - \frac{\Omega_J}{2} -\hat{n}_{\downarrow\uparrow^{\prime\prime},\downarrow\uparrow^{\prime\prime}} \\
\end{array}\right],
\end{eqnarray}
where $\Omega_J=\sqrt{J^2+\delta E_Z^2}$,
$\hat{n}_{\uparrow\downarrow^{\prime\prime},\uparrow\downarrow^{\prime\prime}}= (1+J/\Omega_J)\hat{n}_{}^\prime/2$, $\hat{n}_{\uparrow\downarrow^{\prime\prime},\downarrow\uparrow^{\prime\prime}}=\hat{n}_{\downarrow\uparrow^{\prime\prime},\uparrow\downarrow^{\prime\prime}}=(\delta E_Z/\Omega_J)\hat{n}_{}^\prime /2$, $\hat{n}_{\downarrow\uparrow^{\prime\prime},\downarrow\uparrow^{\prime\prime}}= (1-J/\Omega_J)\hat{n}_{}^\prime/2$. Based on the effective Hamiltonian, the dephasing of a two-qubit CPHASE gate can be studied.
Suppose the control qubit is spin-down, then, it is the relative noise $\hat{n}_{\downarrow\uparrow^{\prime\prime},\downarrow\uparrow^{\prime\prime}}$ between the states $\ket{\downarrow\uparrow^{\prime\prime}}$ and $\ket{\downarrow\downarrow}$ determines the dephasing of a CPHASE gate.

\begin{figure}[]
\centering
\includegraphics[scale=0.5]{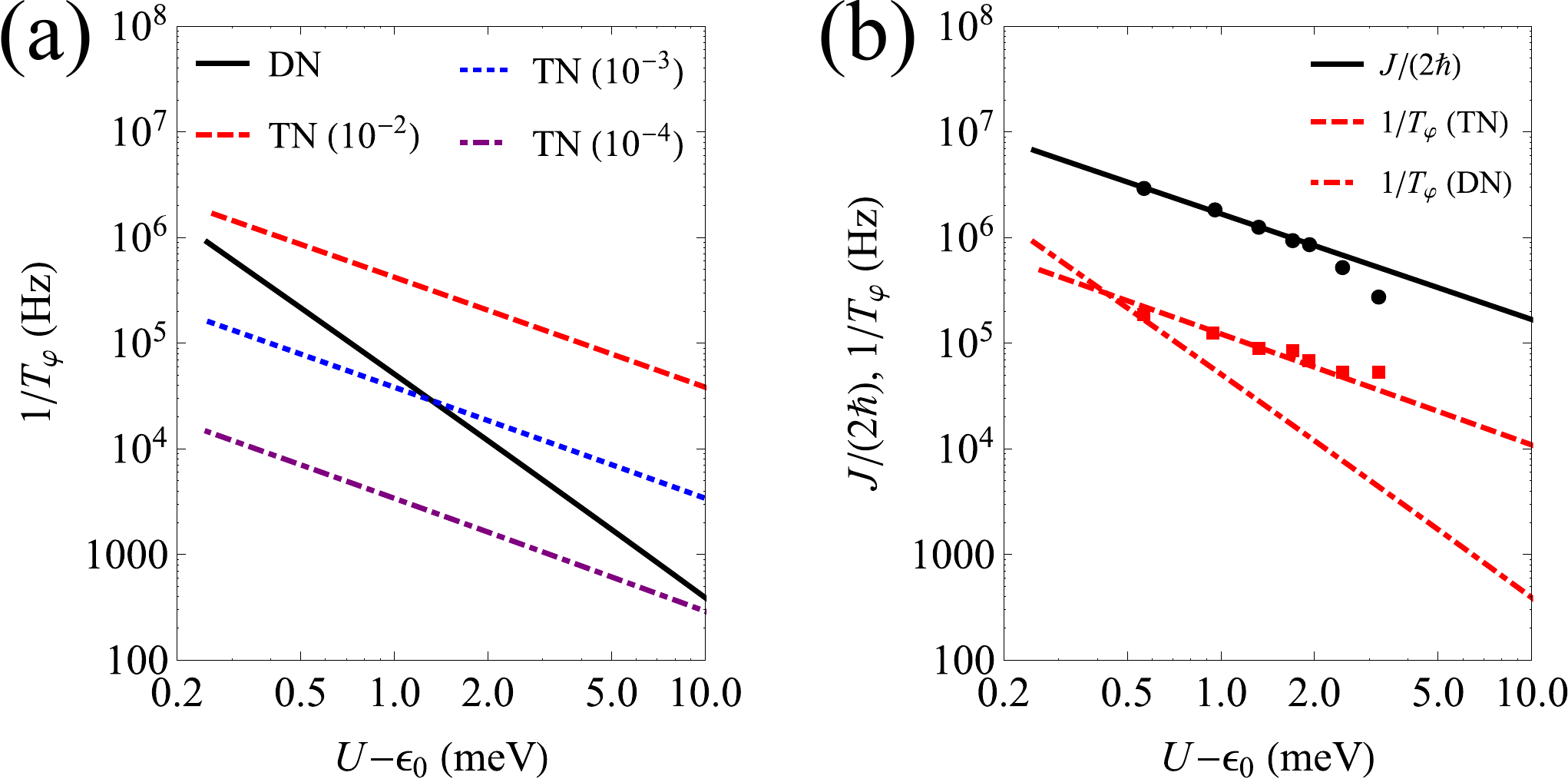}
\caption{Spin dephasing $1/T_\varphi$ as a function of detuning \cite{huang_spin_2018}. (a):
$1/T_\varphi$ as a function of detuning $\epsilon_0$ due to tunneling noise (TN) only or detuning noise (DN) only. For tunneling noise, we choose representative values $\partial t_0/\partial E_b=10^{-2}$ (Dashed), $10^{-3}$ (Dotted), and $10^{-4}$ (Dashdotted).
(b): $J/(2\hbar)$ and $1/T_\varphi$ as a function of detuning $\epsilon_0$, where $A=(2 \mu eV)^2$ and $\partial t_0/\partial E_b=3.2 \times10^{-3}$.  {The dots are the experimental data for $J/(2\hbar)$ (circle) and $1/T_\varphi$ (square) \cite{veldhorst2015}.} }\label{fig:JTphi_detuning}
\end{figure}

Figure \ref{fig:JTphi_detuning}(a) shows the dephasing rate of a CPHASE gate as a function of the detuning in the presence of detuning noise or tunneling noise (the control-qubit is spin-down). Different values of $\partial t_0/\partial E_b$ are chosen to show the possible range of dephasing due to tunneling noise. 
The dephasing due to the detuning noise or the tunneling noise shows different detuning dependences, enables the identification of noise sources. 
%
%
Figure \ref{fig:JTphi_detuning}(b) shows a log-log plot of $J/(2\hbar)$ and dephasing rate $1/T_\varphi$ as a function of detuning in the presence of tunneling noise or detuning noise. The dots are the experimental data from Ref. \cite{veldhorst2015}. 
Note that 
$J/(2\hbar)$ and $1/T_\varphi$ are almost parallel, indicating that they show the same $1/(U-\epsilon_0)$ dependence, which suggests that tunneling noise could dominate spin dephasing in a CPHASE gate. 
%
%
The dominance of tunneling noise happens when $\Delta_b$ is small compared to $U-\epsilon_0$ \cite{huang_spin_2018}, which tends to be satisfied in small accumulation-mode QDs used in the two-qubit experiment \cite{veldhorst2015}.


The number of gate operations also behaves differently for detuning and tunneling noise.
When detuning noise is dominant, the number of CZ operations reduces as the DQD becomes more asymmetric, consistent with experiments in a GaAs DQD \cite{bertrand_quantum_2015, reed_reduced_2016, martins_noise_2016}.
However, when tunneling noise is dominant, the number of CZ operations can increase as DQD becomes more asymmetric (i.e., $\epsilon_0$ approaches ${U}$), as shown in the experiment in silicon \cite{veldhorst2015}. 


\subsection{Suppressing the dephasing in a two-qubit gate}

Here we discuss the mechanisms to suppress dephasing in a two-qubit gate. For the two methods discussed for an $S-T_0$ qubit, only the symmetric operation could in principle suppress the dephasing in a two-qubit gate when detuning noise dominates due to the difference between a two-qubit gate system and an $S-T_0$ qubit in terms of dephasing.
%

The difference between a two-qubit gate and an $S-T_0$ qubit is due to the fact that spin dephasing depends on the relative noise of two states rather than the noise of each individual state.
For an $S-T_0$ qubit,  the qubit is encoded in states $\ket{\uparrow\downarrow^{\prime\prime}}$ and $\ket{\downarrow\uparrow^{\prime\prime}}$, the effective noise is $\hat{h}_{\uparrow\downarrow^{\prime\prime},\downarrow\uparrow^{\prime\prime}}^{(z)}=(\hat{n}_{\uparrow\downarrow^{\prime\prime},\uparrow\downarrow^{\prime\prime}} -\hat{n}_{\downarrow\uparrow^{\prime\prime},\downarrow\uparrow^{\prime\prime}})/2=(J/\Omega_J)\hat{n}_{}^\prime/2$. Increasing $\delta E_Z$, which reduces the ratio $J/\Omega_J$, can reduce the effective noise $\hat{h}_{\uparrow\downarrow^{\prime\prime},\downarrow\uparrow^{\prime\prime}}^{(z)}$ and suppresses the decoherence of $S-T_0$ qubit, as shown in a recent experiment \cite{nichol_high_fidelity_2017}. However, in a two-qubit gate system, for a given state of control-qubit, only one of $\ket{\uparrow\downarrow^{\prime\prime}}$ and $\ket{\downarrow\uparrow^{\prime\prime}}$ is involved. The relative noise will be either $\hat{n}_{\uparrow\downarrow^{\prime\prime},\uparrow\downarrow^{\prime\prime}}$ or $\hat{n}_{\downarrow\uparrow^{\prime\prime},\downarrow\uparrow^{\prime\prime}}$, which is not suppressed with increasing $\delta E_Z$. \textit{Therefore, in contrast to an $S-T_0$ qubit, spin decoherence in a two-qubit logic gate is not suppressed by increasing $\delta E_Z$.}

On the other hand, the symmetric operation reduces the sensitivity of the exchange interaction to the detuning, which suppresses the noise on both the states $\ket{\uparrow\downarrow^{\prime\prime}}$ and $\ket{\downarrow\uparrow^{\prime\prime}}$. When detuning noise is dominant, the symmetric operation should protect a two-qubit gate against charge noise. However, in the presence of tunneling noise, the competition between the tunneling noise and detuning noise could modify the optimal operation point of a two-qubit gate.

\section{Discussion and conclusion}

Spin qubit in silicon has been received considerable attention recently due to the long coherence time and matured fabrication technique in the semiconductor industry. 
The exchange-coupled spin qubit system represents an essential building block for the future scaling-up of spin qubits. More investigations will be needed to further improve the quantum gate fidelity in such systems for applications in quantum computing.
For example, tunneling noise could have a significant effect on electron-spin or hole-spin qubits in donor atoms. It could also have a non-negligible impact on logical spin qubits, such as the $S-T_0$ qubit. If the dephasing mediated by the exchange interaction is suppressed, then, it is necessary to understand the additional spin dephasing mediated by the SOC.} 
{For example, charge noise could induce sizable single-spin pure dephasing through a synthetic SOC in the presence of a micromagnet \cite{yoneda2018, struck_low-frequency_2020, huang_impact_2020}.}
Moreover, in silicon, valley states exist due to the conduction band degeneracy and are coupled to spin degree of freedom \cite{yang2013, hao_electron_2014, zimmerman_valley_2017, schoenfield_coherent_2017, salfi_valley_2018, corna_electrically_2018, penthorn_two-axis_2019, borjans_single-spin_2019,  zhang_giant_2020}. 
To further improve the fidelity of two-qubit gates, spin-echo or other quantum control techniques may be used. It is also necessary to extend the study to multiple spins in multiple QDs, which is essential for the future scaling-up of semiconductor qubits.



In conclusion, we report the recent progress on the study of spin decoherence mechanisms in a DQD for high-fidelity quantum gates.
The schemes are discussed for realizing different types of two-qubit gates in gate-defined QDs or donors and different qubit encodings based on the exchange interaction. Then, we report the recent study on spin dephasing of an $S-T_0$ qubit and a two-qubit gate. The methods are discussed for suppressing the spin dephasing in the system. The study of spin manipulation and dephasing of two electrons in a DQD could provide insight into the realization of high-fidelity two-qubit gates for spin-based quantum computing.

\section*{Acknowledgments}

The results presented here are based on selected works of collaboration. P. H. thanks Xuedong Hu, Garnett W. Bryant, Neil M. Zimmerman, and Dimitrie Culcer for fruitful collaborations.
P.H. acknowledges supports from the National Natural Science Foundation of China (No. 11904157) and Guangdong Provincial Key Laboratory (Grant No.2019B121203002).

\section*{Conflict of Interest}

The authors declare no conflict of interest.

\section*{Keywords}

spin qubit, quantum dot, two-qubit gate, dephasing, exchange interaction, quantum computing

\bibliographystyle{plain}
\bibliography{21a1}


\end{document}